\documentstyle[aps,preprint]{revtex}

\begin{document}
\draft

\title{
A Resonant X-ray Scattering Study of Octahedral Tilt Ordering in
LaMnO$_3$ and Pr$_{1-x}$Ca$_x$MnO$_3$}

\author{M. v. Zimmermann, C.S. Nelson, Y.-J. Kim, J.P.
Hill, Doon Gibbs}
\address{Physics Department, Brookhaven National Laboratory, Upton, New
York  11973-5000}
\author{H. Nakao, Y. Wakabayashi*, Y. Murakami}
\address{Photon Factory, Institute of Materials Structure Science,
Tsukuba 305-0801, Japan}
\author{Y. Tomioka$\dag$, Y. Tokura}
\address{Joint Research Center for Atom Technology (JRCAT), Tsukuba
305-0046, Japan}
\author{T. Arima}
\address{Institute of Materials Science, University of Tsukuba,
Tsukuba 305-8573, Japan}
\author{C.-C. Kao}
\address{National Synchrotron Light Source, Brookhaven National
Laboratory, Upton, New York 11973-5000}
\author{D. Casa, C. Venkataraman, Th. Gog}
\address{CMC-CAT, Advanced Photon Source, Argonne National Laboratory,
Argonne, Illinois  60439}

\footnotetext{* Also at Department of Physics, Faculty of Science
and Technology, Keio University, Yokohama 223-8522, Japan}

\footnotetext{$\dag$ Also at Department of Applied Physics,
University of Tokyo, Tokyo 113-8656, Japan}

\maketitle
\newpage
\begin{abstract}
We report an x-ray scattering study of octahedral tilt ordering in
the manganite series Pr$_{1-x}$Ca$_x$MnO$_3$ with x=0.4 and 0.25
and in LaMnO$_3$.  The sensitivity to tilt ordering is achieved by
tuning the incident x-ray energy to the L$_I$, L$_{II}$ and
L$_{III}$ absorption edges of Pr and La, respectively.  The
resulting energy-dependent profiles are characterized by a
dipole-resonant peak and higher energy fine structure.  The
polarization dependence is predominantly $\sigma$-to-$\pi$ and the
azimuthal dependence follows a sin-squared behavior.  These
results are similar to those obtained in recent x-ray scattering
studies of orbital ordering carried out in these same materials at
the Mn K edge. They lead to a description of the cross-section in
terms of Templeton scattering in which the tilt ordering breaks
the symmetry at the rare earth site. The most interesting result
of the present work is our observation that octahedral tilt
ordering persists above the orbital ordering transition
temperatures in all three samples. Indeed, we identify separate
structural transitions which may be associated with the onset of
orbital and tilt ordering, respectively, and characterize the loss
of tilt ordering versus temperature in LaMnO$_3$.

\end{abstract}

\newpage

\section{Introduction}

The AMO$_3$ aristotype is the defining feature of the cubic
perovskite structure.  It consists of a cubic array of
metal-oxygen octahedra centered on a lanthanide ion (A).  In the
transition metal oxides, the aristotype is generally observed only
at high temperatures.  At lower temperatures, a variety of
distortions of the octahedra may occur, which reduce the cubic
symmetry. It is interesting that most, if not all, of the unusual
properties exhibited by these materials, including colossal
magnetoresistance and high temperature superconductivity, occur in
the lower temperature range.  It remains an open question whether
these distortions are required for such phenomena to occur.

A prominent example of an octahedral distortion is the Jahn-Teller
distortion in the manganites. In specific cases, this involves the
lengthening of two Mn-O bonds of the MnO$_6$ octahedra and the
concomitant shortening of four others.  It is accompanied by
orbital ordering of the occupied Mn 3d orbitals.  Likewise, charge
ordering involves the inhomogeneous localization of electronic
charge at Mn sites to form ordered arrays. Both orbital and charge
ordering introduce new periodicities into the system and may be
accompanied by longitudinal or transverse lattice distortions.
Characterizing the charge, orbital and magnetic ordering in
transition metal oxides, and exploring their coupling to the
lattice, is an active field of current research
\cite{Orenstein00,vZim1,Tokura}.

Another important distortion of the cubic perovskites involves the
tilting of the metal-oxygen octahedra. These ordered rotations are
well-known, having been discussed by Goldschmidt in the nineteen
twenties\cite{Goldschmidt26}, and again by
Goodenough\cite{Goodenough55} in the nineteen fifties. A complete
crystallographic treatment of 23 possible tilt patterns has been
given by Glazer \cite{Glazer72}, and refined more recently by
Woodward \cite{Woodward1,Woodward2}. The underlying idea is that a
small tilt of the octahedra shortens the A-site ion-to-oxygen bond
distances, while preserving the length of the metal-oxygen bond --
thereby lowering the total energy.  A three dimensional view of
one common tilt ordering, the so-called GdFeO$_3$ distortion, is
illustrated in Figure 1, together with three two-dimensional
projections.  In this distortion, the MnO$_6$ octahedra are
rotated by $\sim \pm$ 9$^{\circ}$ about the b-axis and by a small
angle ($< 1^\circ$) about the a-axis using the $\it{Pbnm}$ setting
\cite{Glazer}. It is intriguing that in LaMnO$_3$ the periodicity
of the Mn orbital ordering is the same as that of the octahedral
tilt ordering. This raises the questions of whether the tilt
ordering is coupled to the charge and orbital ordering, possibly
as a precursor, and whether the tilting plays a more important
role in the properties of CMR materials than has been recognized
to date. In a recent paper, Mizokawa et al.\cite{Mizokawa99}
addressed these questions in LaMnO$_3$ and Pr$_{1-x}$Ca$_x$MnO$_3$
(hereafter referred to as PCMO), among others.  They showed that
the GdFeO$_3$ distortion stabilizes the so-called d-type orbital
ordering pattern in LaMnO$_3$ by increasing the hybridization of
the La atoms with three of the corner oxygen atoms. However,
experimental studies of these effects in manganites have been
limited\cite{Rodriguez98}.

In this paper we discuss x-ray resonant scattering studies of the
octahedral tilt ordering observed in LaMnO$_3$,
Pr$_{0.75}$Ca$_{0.25}$MnO$_3$ and Pr$_{0.6}$Ca$_{0.4}$MnO$_3$. In
analogy to recent studies of orbital ordering in which the
incident energy is tuned to the Mn K
edge\cite{Murakami214,Ishihara98,vZim1}, we show that in each case
it is possible to probe octahedral tilt ordering by tuning the
photon energy to the respective La and Pr L absorption edges.
Specifically, we report resonant spectra of the tilt ordering at
the La and Pr L$_I$ (2s $\rightarrow$ 6p), and L$_{II-III}$ (2p
$\rightarrow$ 5d) edges, including the azimuthal dependence at the
Pr L$_{II}$ edge of Pr$_{0.75}$Ca$_{0.25}$MnO$_3$. We find that
the resonant fine structures at the L$_{II}$ and L$_{III}$ edges
are qualitatively similar, involving a main resonant peak and
weaker subsidiary peaks, whereas the L$_I$ spectra involve only a
single resonance. We believe these differences reflect the
electronic structure of the intermediate states, however, a full
theoretical analysis including the 5d and 6p bands is required
before quantitative conclusions can be drawn. The polarization
dependence of the resonant scattering is predominantly $\sigma
\rightarrow \pi$ to within experimental errors, and the azimuthal
dependence follows a sin-squared behavior. Both are consistent
with simple models of Templeton scattering based on a splitting of
the resonant intermediate
states\cite{Templetons,vZim1,Murakami113,Ishihara98}. Where
possible, specific comparisons are made with the predictions of
Benedetti \textit{et al.}\cite{Benedetti}, who have carried out
LDA + U calculations  of these effects for LaMnO$_3$. No evidence
has been found for lattice modulations originating with a
longitudinal periodic displacement of the A-site ions away from
their symmetry positions in LaMnO$_3$ and
Pr$_{0.75}$Ca$_{0.25}$MnO$_3$, however, the possibility of a
transverse displacement has not yet been explored.

The most interesting results of the present work concern the
temperature dependence of the octahedral tilt ordering. In
particular, we find that the tilt ordering in LaMnO$_3$ persists
above the orbital ordering transition at $\sim$ 800 K and then
disappears at $\sim$ 1000 K.  Both the tilt and orbital ordering
transitions are associated with corresponding structural
transitions of the lattice, which we specify. Interestingly, the
line width of the octahedral tilt ordering is resolution limited
below the orbital ordering temperature, but starts to increase
about 70 K above it. No evolution of the main peak of the resonant
lineshapes of the L$_{II}$ and L$_I$ absorption edges was found
for temperatures increasing above the orbital ordering transition,
which differs from the predictions of Benedetti, \textit{et
al.}\cite{Benedetti}.

The situation is more complicated in Pr$_{0.75}$Ca$_{0.25}$MnO$_6$
for which the structure of the orbital ordering is itself
controversial. Although most of the orbital order scattering in
this material disappears at about 400 K, coincident with a
structural transition, a small peak nevertheless persists at the
Mn K edge up to 900 K. In contrast, the intensity of the resonant
tilt scattering measured at the Pr L$_{II}$ edge is approximately
constant from 10-900 K, the highest temperature accessible with
the present apparatus.  Similarly, there is little change of the
resonant lineshape at the L$_I$ and L$_{II}$ absorption edges of
Pr$_{0.6}$Ca$_{0.4}$MnO$_3$, as the temperature is increased
through its orbital ordering temperature (T$_{co} \sim$ 240 K),
nor are there variations of the tilt scattering intensity,
correlation length or periodicity, at least up to about 300 K.

Taken together, these results offer a new probe of octahedral tilt
ordering and of its possible coupling to charge and orbital
ordering in the pseudo-cubic perovskites. In the following, we
briefly describe the experimental set-up and the phase behavior of
LaMnO$_3$ and Pr$_{1-x}$Ca$_x$MnO$_3$ before turning to a
discussion of our results.

\section{Experimental}

The single crystals used in the present experiments were grown by
floating zone techniques at JRCAT.  (1,0,0) and (0,1,0) surfaces
were cut from cylinders of radius $\sim$3 mm, and polished with
fine emery paper and diamond paste.  The mosaic widths of the
LaMnO$_3$ and PCMO samples, as characterized at the (0,2,0) bulk
Bragg reflections (in orthorhombic notation), were all about
0.1$^o$ (FWHM), as described
earlier\cite{vZim1,Murakami113,Zimmermann99}. These values varied
by small amounts as the beam was moved across each sample surface,
reflecting its mosaic distribution. The growth techniques and
basic transport properties have been described in detail
elsewhere\cite{Okimoto98,Tomioka96,Tokura99}.

X-ray scattering experiments were carried out at the National
Synchrotron Light Source on bending magnet X22C and wiggler X21
beamlines, and at the Advanced Photon Source on beamline 9ID. X22C
is equipped with a bent, toroidal focusing mirror and a Ge(1,1,1)
double crystal monochromator arranged in a vertical scattering
geometry. This gives an incident linear polarization of 95$\%$
($\sigma$) and an incident energy resolution of between 5 and 10
eV at the Mn K and La and Pr L$_I$, L$_{II}$ and L$_{III}$ edges.
Two different detector configurations were used. High momentum
transfer resolution scans employed a Ge(1,1,1) crystal, and gave a
longitudinal resolution of 4.5 x $10^{-4}$\AA$^{-1}$(HWHM) at the
respective (0,1,0) reflections and linear polarization analysis of
the scattered beam was provided via rotation of a Cu(2,2,0)
crystal around the scattered beam direction\cite{Gibbs88}.  The
latter gave longitudinal resolutions of $0.0069$\AA$^{-1}$ and
$0.0052$\AA$^{-1}$ (HWHM) in the $\sigma \rightarrow \sigma$ and
$\sigma \rightarrow \pi$ geometries, respectively.  NSLS Wiggler
beamline X21 was equipped with a 4-bounce Si(2,2,0) monochromator
and a focusing mirror, leading to an incident energy resolution of
0.25 eV. APS undulator beamline 9ID consists of a double crystal
Si(1,1,1) monochromator, a focussing mirror (coated with Pt) and a
flat harmonic rejection mirror.  The incident energy resolution
was approximately 1.5 eV.

A serious complication associated with the present experiments was
our observation that these crystals do not tolerate repeated
excursions to high temperatures \( > 800 K\) before changing their
properties.  The most dramatic of the effects we observed was the
loss of long-range orbital order in one of the LaMnO$_3$ samples
after extended cycling above 1000K, and cooling back to room
temperature.  These effects are common among perovskites and
originate in strain relief within individual grains and, perhaps,
in oxygen depletion\cite{Vogt,Granado00}. As a consequence, our
ability to double check all of the results obtained for LaMnO$_3$
has been limited.

At low temperatures, Pr$_{0.75}$Ca$_{0.25}$MnO$_3$ and LaMnO$_3$
have pseudo-cubic perovskite structures with orthorhombic symmetry
$\it{Pbnm}$ (illustrated in Figure 1). Each Mn atom lies at the
center of an octahedron defined by six oxygen atoms at the
corners.  Single layers of La or Pr/Ca atoms lie between the
layers of octahedra. Both materials are insulating and believed to
exhibit an orbitally ordered ground state. The electronic
configurations of the (Mn)$^{3+}$ (d$^4$) ions are ($t^3_{2g}$,
$e^1_g$) with the $t_{2g}$ electrons localized at the Mn sites.
The e$_g$ electrons are hybridized with the oxygen 2p orbitals,
and participate in a cooperative Jahn-Teller distortion of the
MnO$_6$ octahedra.  This is believed to lead to a
($3x^2-y^2$)-($3y^2-r^2$) zig zag-type of orbital ordering of the
e$_g$ electrons with the oxygen displaced along the direction of
extension of the e$_g$ orbitals, as has been discussed in detail
elsewhere\cite{Orenstein00,vZim1,Goodenough55}.  In orthorhombic
notation, for which the fundamental Bragg peaks occur at (0,2k,0),
with k an integer, the orbital scattering occurs at (0,2k+1,0).
Regarding their magnetic structures, Pr$_{0.75}$Ca$_{0.25}$MnO$_6$
is ferromagnetic, while LaMnO$_3$ is an A-type antiferromagnet,
below their respective magnetic ordering temperatures.

Pr$_{0.6}$Ca$_{0.4}$MnO$_3$ also has a perovskite structure with
$\it {Pbnm}$ symmetry, but exhibits an insulating CE-type,
antiferromagnetic ground state at low temperature, including a
cooperative Jahn-Teller distortion. The higher Ca doping in this
material leads to a large negative magnetoresistance in an applied
magnetic field with the metal-insulator transition occurring at
about 6T at low temperature\cite{Tomioka96}. The ordered phase is
accompanied by a second modulation of the lattice arising from
charge ordering among Mn$^{3+}$ and Mn$^{4+}$ ions, which occurs
in addition to orbital ordering\cite{vZim1}. The large
conductivity is enabled through the motion of electrons formerly
localized at (Mn)$^{3+}$ sites. In orthorhombic notation, the
charge order reflections in Pr$_{0.6}$Ca$_{0.4}$MnO$_3$ occur at
(0,2k+1,0), whereas the orbital reflections are at (0,k+1/2,0).
Note that the orbital period present in the x=0.4 sample (=2b)
differs from that in the x=0.25 sample (=b) as a result of the
charge ordering. Detailed x-ray scattering studies of the charge
and orbital ordering in LaMnO$_3$ and PCMO have been reported
elsewhere \cite{vZim1,Murakami214,Murakami113,Zimmermann99}.

In addition to the Jahn-Teller distortion, all three compounds
undergo the so-called GdFeO$_3$ distortion, which involves tilting
of the MnO$_6$ octahedra, as shown in Figure 1. In the GdFeO$_3$
distortion, the four octahedra in the unit cell are rotated by an
angle $\omega$ around an axis in the (0,1,1) plane. This may be
approximated as a compound rotation of the octahedra first about
the b-axis (fig. 1d) followed by a much smaller rotation about the
a-axis (fig. 1c).  From the present perspective, it is important
to note that the periodicity of the octahedral tilt ordering in
LaMnO$_3$ and Pr$_{0.75}$Ca$_{0.25}$MnO$_3$ is identical to that
of the orbital ordering (specified above). In contrast, the
octahedral tilt ordering in Pr$_{0.6}$Ca$_{0.4}$MnO$_3$ has the
same period as the charge ordering, which is half that of the
orbital ordering. Crystallographic studies of the structure of
LaMnO$_3$ \cite{Rodriguez98,Huang} suggest that the octahedral
tilting angle is reduced from 16 to 12$^o$, but not eliminated, by
increasing the sample temperature from 300 to 800 K.  They also
suggest that both the oxygen atoms and A-site ions are displaced
by octahedral tilting, all the while maintaining $\it{Pbnm}$
symmetry.  To our knowledge, there are no comparable
crystallographic studies of PCMO for the dopings considered here,
but qualitative similarities are expected.

\section{Results and Discussions}
\subsection{Resonant Profiles}

Figure 2 shows the energy dependence of the scattering at the
(0,1,0) orbital wavevector of Pr$_{0.75}$Ca$_{0.25}$MnO$_3$ as the
incident x-ray energy is tuned through the Mn K and Pr L$_{II}$
absorption edges. These data were obtained using a polarization
analyzer and explicitly resolve the $\pi$-component of the
resonant scattering, consistent with a rotation of the incident
linear polarization from $\sigma$ to $\pi$.  Referring to the
figure, a large resonant signal is visible at {$\hbar\omega$=6.547
keV}, reaching about 500 counts/sec near the Mn K edge. In
addition, there are two smaller peaks at $\hbar\omega$=6.56 and
6.575 keV and a broad peak at $\hbar\omega$=6.62 keV. Remarkably,
there is also a resonant feature at the Pr L$_{II}$ edge with
$\hbar\omega$=6.44 keV.  It is this latter resonance which is the
subject of the present paper.

Detailed scans of the resonant scattering obtained at the orbital
wavevectors of Pr$_{0.75}$Ca$_{0.25}$MnO$_3$ and LaMnO$_3$ for
incident x-ray energies tuned through the L$_I$, L$_{II}$ and
L$_{III}$ edges of Pr and La are shown in Figures 3 and 4,
respectively. These data were obtained at the (0,1,0) reflections
of each sample using a Ge(1,1,1) analyzer crystal, and thereby
combine any $\sigma$ and $\pi$ components of the scattering.
Similar results for Pr$_{0.6}$Ca$_{0.4}$MnO$_3$
(polarization-resolved) are shown in Figure 6, and discussed
later.

The data for the L$_I$ edges are simplest, consisting in each case
of a resonant profile centered near the white line of the
fluorescence, with a width of about 10 eV. Fits of the lineshapes
to squared Lorentzians suggest the possibility of a weak asymmetry
with additional intensity at lower energy. Both samples also show
additional, broad scattering peaked between $\sim$ 50 and 80 eV
above the L$_I$ edge (not shown).  Explicit tests of these latter
peaks for multiple scattering were carried out for both PCMO
samples by repeating the energy scans at several different
azimuths. These tests showed that the broad peaks above the edge
arise from multiple scattering.  We suspect that the same is true
for the LaMnO$_3$ but have not yet carried out such explicit
tests.

The profiles of the Pr scattering at the L$_{II-III}$ edges show
more fine structure (Fig. 3). In each case, the main resonance
occurs 2-3 eV above the inflection point of the fluorescence line,
and has a full width of between 6 and 9 eV\cite{fullwidths}. Fits
of the main resonance peaks to squared Lorentzians again suggest
that the lineshapes are not symmetric, in this case having longer
high energy tails.  It is not clear from the data whether this
asymmetry reflects an additional excitation channel (simply adding
to the intensity) or instead arises from an interference effect,
or from multiple scattering. In addition to the main resonance,
there is a second peak occurring about 30 eV above the resonance
at both the L$_{II}$ and L$_{III}$ absorption edges. We have ruled
out multiple scattering as the origin of these latter peaks and
believe that they are of physical interest. Polarization analysis
reveals that both the main resonant peak and the second peak 30 eV
above it, occur predominantly in the $\sigma-\pi$ channel. More
specifically, we find that the main resonant feature at the Pr
L$_{II}$ edge of Pr$_{0.6}$Ca$_{0.4}$MnO$_3$ gives a ratio of the
$\sigma-\pi$/$\sigma-\sigma$ scattering of at least 25. Similar
results have been obtained for the x=0.25 sample.

The resonant lineshapes obtained at the La L$_{II}$ and L$_{III}$
edges in LaMnO$_3$ are qualitatively similar to those obtained at
the Pr L edges, and are shown in Figure 4. In each case, the main
resonant peak occurs 2-3 eV above the inflection point of the
fluorescence, and exhibits a slight asymmetry to higher energy.
There are additional peaks occurring about 30 eV above the main
resonance, similar to those observed in the PCMO samples. As we
have not yet adequately characterized the multiple scattering in
this sample, we cannot rule it out at this second peak position in
LaMnO$_3$.  Given the similarity of these features with those
observed in PCMO, where multiple scattering was ruled out, it is
tempting to conclude that it is not a factor here. We suspect,
however, that the shoulder located about 10 eV below the L$_{II}$
absorption edges (see fig. 4) and a broad peak centered at 5.405
keV (whose tail is visible at low energy in the scan of the
L$_{III}$ absorption edge) do arise from multiple scattering.
These latter details will have to be confirmed in future
measurements, however, their outcome will not detract from the
main results presented here.

An important feature of the resonant scattering is the dependence
of the intensity on azimuthal angle.  The azimuthal angle $\psi$
characterizes rotations of the sample about the scattering
wavevector and is defined to be zero when the c-axis is
perpendicular to the scattering plane. A quantitative study of the
azimuthal dependence of the L$_{II}$ resonant scattering of
Pr$_{0.75}$Ca$_{0.25}$MnO$_3$ is shown in Figure 5. Each data
point represents the maximum resonant intensity obtained in the
$\sigma-\pi$ geometry at the (0,1,0) reflection at a particular
azimuthal angle. The data have been normalized by the intensity of
the (0,2,0) reflection at that azimuth to correct for small
variations due to sample shape. In contrast to normal charge
scattering, for which the intensity is independent of the
azimuthal angle, the resonant scattering exhibits an oscillation
with two-fold symmetry. The intensity approaches zero when
$\psi$=0 and 180$^{\circ}$, similar to the $\sigma-\pi$ polarized
component of the orbital scattering measured at the Mn K edge. The
solid line is a fit to the form $A\sin^2\psi$, as was also found
for the azimuthal dependence of the orbital ordering at the Mn K
edge\cite{vZim1}.

Before discussing the Pr scattering in more detail, we first
discuss the main peak and fine structure near the Mn K edge (shown
in Figure 2). These features have been observed previously and
interpreted in terms of orbital ordering \cite{vZim1}. Briefly,
two kinds of Mn site are distinguishable depending on the local
orientation of the e$_g$ electron, as a result of the cooperative
Jahn-Teller distortion. The resonant scattering observed at the
orbital wavevector may then be thought of as Templeton scattering
arising from the anisotropic charge distribution induced by
orbital ordering\cite{Templetons,Blume96}. In the dipole
approximation, the resonance corresponds to a 1s $\rightarrow$ 4p
transition at the metal site. In the simplest model, the
sensitivity to orbital ordering arises from the splitting of the
Mn 4p levels as a result of the 3d ordering. Discussion of the
microscopic origin of the splitting, however, has been
controversial\cite{Ishihara98,Fabrizio98,Elfimov99,Benfatto99,IshiharaPRB,vZim1,Benedetti,Takahashi}
with both Coulomb- and Jahn-Teller- based descriptions proposed.
Insofar as we are aware, the experimental data obtained to date do
not distinguish either approach conclusively, and this remains an
open question.

Regardless of the origin of the splitting, the resonant scattering
at the Mn K edge reflects the symmetry of the orbital ordering
through the redistribution of local charge density (and subsequent
perturbation of the electron energy levels) at the Mn$^{3+}$
sites, consistent with the space group. It follows that the peak
positions and widths measure the orbital periodicity and
correlation lengths, respectively.  A key feature of this type of
resonant scattering at an (0,k,0) reflection in LaMnO$_3$ and PCMO
is that the azimuthal dependence of the intensities for rotations
of the sample about the momentum transfer follows a $\sin^2\psi$
dependence.  In addition, the polarization of the incident beam is
rotated from $\sigma$ to $\pi$. Full details of these effects may
be found elsewhere, including a generalization to the orbital
ordering of the t$_{2g}$ levels and to other geometries
\cite{vZim1,Murakami214,Ishihara98,Murakami113,IshiharaPRB}.

Returning to the Pr resonance, there are at least two possible
sources of this scattering.  One is that it results from orbital
ordering of the partially occupied Pr 4f levels, with a
periodicity equal to the Mn 3d orbital order.  Such
antiferro-quadrupolar ordering has recently been observed in the
Dy 4f levels of DyB$_2$C$_2$ at the Dy L$_{III}$
edge\cite{Hirota,Hirota2}. However, this cannot be the origin of
these effects in LaMnO$_3$, since La lacks 4f electrons. A second
possibility is that it originates in the differing kinds of Pr
environment that exist within the unit cell. While all Pr sites
are crystallographically equivalent (resulting in the (0,1,0)
being forbidden in $\it{Pbnm}$ symmetry), there are in fact two Pr
environments created by the surrounding oxygens in the tilted
phase.  These are simply illustrated in Fig. 1. Conventional x-ray
scattering is insensitive to such differences, however, at an
absorption edge the Pr scattering factor transforms from a scalar
to a tensor reflecting the anisotropies of the resonant
ion\cite{Templetons,Blume96}. The tensors at the two Pr sites are
inequivalent, and the (0,1,0) reflection becomes allowed, thereby
giving rise to Templeton scattering.  Since these two environments
result from the tilting of the oxygen octahedra, it follows that
the resonant intensity is a measure of the octahedral tilt
ordering, with the width of the scattering in reciprocal space
providing information on the correlation length of the tilt
ordering, and its position giving the periodicity.

We, therefore, interpret the resonant scattering at the La and Pr
L edges discussed above as Templeton scattering associated with
the anisotropic charge distribution induced by octahedral tilt
ordering, and consistent with the $\it{Pbnm}$ space group. The
energies of the resonances in each case suggest dipole excitations
coupling 2p$_{3/2} \rightarrow $5d$_{3/2,1/2}$, 2p$_{1/2}
\rightarrow$ 5d$_{3/2}$, and 2s $\rightarrow$ 6p states for the
L$_{III}$, L$_{II}$ and L$_{I}$ absorption edges, respectively. We
speculate that the splitting $\delta_o$ in the intermediate states
needed for the resonant cross-section to give a non-zero intensity
originates in the crystal field splitting of the appropriate
states at the Pr and La sites following tilting of the octahedra.
Given a splitting $\delta_{o}$, the same arguments which were used
to describe the Mn K edge resonant scattering at the orbital
wavevector may then be applied to describe the scattering from the
tilt ordering, and naturally lead to both a sin-squared dependence
on the azimuthal angle and a rotated $(\sigma-\pi)$ final
polarization, as observed.  The differences between the resonant
L$_{II-III}$ and L$_I$ lineshapes--in particular, the apparent
absence of higher energy fine structure at the L$_I$
edge--probably reflect the differing density of states of the 5d
and 6p intermediate states, respectively. Clarifying this point
will require more detailed band-structure calculations.

It is interesting to compare these results with the predictions of
Benedetti, \textit{et al.}\cite{Benedetti}, who have carried out
LDA+U calculations of the resonant scattering expected at the La
edges of LaMnO$_3$ in the presence of both orbital and tilt
ordering. The main features observed in our data are predicted,
including a dipole resonance, a rotated ($\sigma-\pi$)
polarization dependence and sin-squared azimuthal dependence.  Our
x-ray results also show subsidiary peaks at higher energy, which
are qualitatively consistent with the predictions, although the
observed peaks fall $\sim$ 10 eV higher than is predicted. These
calculations support our identification of the resonant scattering
at the La and Pr edges as being associated with the tilt ordering.

A fascinating question concerns the possibility that the orbital
ordering of the Mn e$_g$ electrons (and concomitant Jahn-Teller
distortion) might also contribute to the resonant scattering at
the La and Pr edges through hybridization of the oxygen 2p states
with the 5d, 6p, and 3d states of the La/Pr and Mn, respectively.
Indeed, the work of Benedetti, \textit{et al.} on
LaMnO$_3$\cite{Benedetti} suggests that the resonant profile
calculated by including only the effects of tilt ordering shifts
to slightly lower energy when orbital ordering is subsequently
included.  We have tested this prediction in
Pr$_{0.6}$Ca$_{0.4}$MnO$_3$ and LaMnO$_3$ by measuring the
resonant profiles of the tilt ordering at the Pr L$_I$, L$_{II}$
edges and at the La L$_I$ edge for temperatures above and below
their respective orbital ordering temperatures. Figures 6a and b
show the energy dependence of the resonant scattering of
Pr$_{0.6}$Ca$_{0.4}$MnO$_3$ obtained at the Pr L$_I$ and L$_{II}$
absorption edges.  The data were measured at the (0,1,0)
reflection, which for this doping is coincident with a charge
order reflection.  In order to discriminate against the
contribution of the charge ordering, which is predominantly
$\sigma-\sigma$ polarized\cite{vZim1}, these data were obtained in
a $\sigma-\pi$ geometry using the polarization analyzer. Open
circles show the results obtained below the orbital ordering
temperature at 200 K and closed circles show the data obtained
above at 280 K.  (T$_{oo}$ = 245 K in this sample.)  The main
features of the resonant profiles discussed above for
Pr$_{0.75}$Ca$_{0.25}$MnO$_3$, including the slight asymmetries
and additional fine structure are reproduced at the higher
temperature.  More importantly, from the perspective of the theory
of Benedetti, \textit{et al.}\cite{Benedetti}, no changes in the
L-edge lineshapes are observed with increasing temperature through
the orbital ordering transition.  From this we conclude that the
main features of the L-edge resonant profiles reported here
originate in octahedral tilt ordering, and that any contributions
to the L edge resonances arising from the Mn orbital order, e.g.,
from changes of the La/Pr hybridization with the oxygen motion,
are too small to be detected, at least at the present experimental
signal rates.

Qualitatively similar results were also obtained at the La L$_I$
edge of LaMnO$_3$ for temperatures above and below its orbital
ordering temperature.  A surprising additional result in this
sample, however, involved the observation of a continuous shift of
the second peak at about 30 eV above the main resonance with
increasing temperature. In particular, for temperatures increasing
from about 650 K to 800 K, this second resonant peak increased its
position by $\sim$ 15 eV. We suspect that multiple scattering or
some other artifact, such as oxygen depletion, may be the origin
of this effect, however, more systematic tests will require a new
sample.

\subsection{Temperature Dependence}

The temperature dependence of the resonant intensity of the
octahedral tilt ordering obtained at the La L$_I$ absorption edge
of LaMnO$_3$ is shown by the open circles in Figure 7a. For
comparison, filled circles show the temperature dependence of the
resonant intensity of the orbital ordering obtained at the Mn K
edge, which is qualitatively consistent with earlier published
results\cite{Murakami113}.  Each data point represents the peak
intensity taken from a scan through the (1,0,0) reflection
obtained in a high-resolution mode using a Ge(1,1,1) analyzer. All
of the data were obtained using a high temperature oven. Referring
to the figure, the La L$_I$ and Mn K resonant intensities are both
approximately constant between 300 and 700 K, above which
temperature both begin to decrease. The intensities at the La
L$_I$ and Mn K edges then fall abruptly to near-zero at about 800
K, which corresponds to an orthorhombic-to-orthorhombic structural
phase transition (O $\rightarrow$ O').  No significant
Mn-resonant scattering was observed at the (1,0,0) position of the
low temperature orthorhombic structure (O) at any temperature
above 800 K, although a weak, broad peak, consistent with orbital
fluctuations, was observed. The intensity of this diffuse
scattering falls gradually with increasing temperature and
disappears above 1000 K. A strong La-resonant peak re-appears
above 800 K, but is located at the (1,0,0) position of the high
temperature orthorhombic phase O' (filled circles). The
intensity of the La-resonant scattering in the high temperature
phase $Ó$ increases with temperature until about 870 K, and then
decreases again until it disappears at about 1000 K. This suggests
that the tilt ordering is preserved within the new lattice,
consistent with the results of powder diffraction studies
\cite{Rodriguez98} (which also report a reduced average tilt
angle).  It follows that the temperature dependences of the tilt
and orbital ordering are distinctly different, with the octahedral
tilting persisting to higher temperatures.

The temperature dependences of the corresponding half-widths at
half-maximum (HWHM) are plotted on a log-scale for the Mn- and
La-resonant scattering in Fig. 7b.  These are proportional to the
inverse correlation lengths of the Mn orbital and tilt order,
respectively.  For temperatures below about 800 K, the widths of
both the orbital and tilt scattering are consistent with the
resolution, implying correlation lengths of 1000 $\AA$, or
greater. Above 800 K, the half width of the orbital scattering
abruptly broadens by a factor of ten (see Fig. 7b), corresponding
to correlation lengths of about $\sim$ 100$\AA$. The half width of
the tilt scattering, however, remains constant up to about 870 K,
at which temperature the tilt intensity begins to decrease. Above
this temperature, the width increases continuously. No significant
broadening was observed at the (200) bulk Bragg peak over the
entire temperature range.

On the basis of these results, we associate the
orthorhombic-to-orthorhombic transition at 800 K with the
destruction of long ranged orbital ordering in LaMnO$_3$, in
agreement with earlier conclusions\cite{Murakami113,Rodriguez98}.
Explicit measurements of the bulk (2,0,0) intensities and
positions versus temperature reveal the collapse of the a and b
lattice constants to nearly equal values at this temperature,
consistent with the appearance of a pseudocubic
phase\cite{Rodriguez98}.  We associate the broad scattering which
occurs at the orbital wavevector above 800 K with critical orbital
fluctuations, such as have been observed earlier in PCMO for x=0.4
and 0.5\cite{Zimmermann99,vZim1}.  It is interesting that the
full-widths do not continuously broaden above 800 K, but remain
approximately constant, reminiscent of studies of polaron
correlations in PCMO and LCMO\cite{Nelson}. Accompanying the loss
of orbital ordering at 800 K is the apparent loss of octahedral
tilt ordering in the low temperature O phase, however, the tilt
ordering re-appears in the high temperature O' phase with
resolution-limited line widths. It persists until about 870 K,
when the intensities and the correlation lengths start to
decrease. Octahedral tilt ordering disappears at $\sim$ 1000 K,
corresponding to a structural transition to a rhombohedral
phase\cite{Rodriguez98}.  The fact that the octahedral tilting
achieves long-range order at temperatures above the orbital
ordering transition appears consistent with the conclusion of
Mizokawa, \textit{et al.}\cite{Mizokawa99} that the GdFeO$_3$
distortion stabilizes one type of orbital ordering in LaMnO$_3$ in
preference to another, and suggests that octahedral tilt ordering
may serve as a precursor to orbital ordering.  It raises the
general question of the nature of the coupling between the tilt
ordering and the other degrees of freedom in these
systems--especially between the orbital and tilt ordering.

The temperature dependence of the intensities of the tilt ordering
in Pr$_{0.75}$Ca$_{0.25}$MnO$_3$, as obtained at the Pr L$_{II}$
edge, is shown by the filled circles in Fig. 8a. Open circles
represent the temperature dependence of the intensity of the
orbital scattering obtained at the Mn K edge, and have been
published previously\cite{vZim1}.  As above, each data point
corresponds to the peak intensity resulting from a scan through
the (0,1,0) reflection, which is the orbital wavevector in this
sample.  The data were obtained using a displex at low temperature
and an oven at high temperatures, and then scaled to be equal at
300 K. As may be seen from the figure, the Pr-resonant intensities
are approximately constant over the entire range accessible
between 10 and 900 K. In contrast, the temperature dependence of
the orbital scattering as measured at the K-edge is approximately
constant between 10 and 200 K, and then decreases between 200 and
400 K with a long tail extending up to about 850 K. High-momentum
transfer resolution measurements made at both the Pr L$_{II}$ and
the Mn K absorption edges show that the peak widths of the Pr- and
Mn- scattering remain approximately constant below 900 K,
corresponding to correlation lengths of $\sim$ 2000 $\AA$, or
greater, in each case (see Figure 8b). It is natural to associate
the decrease of the resonant orbital scattering between 200 and
400 K with the orthorhombic structural transition reported by
Jirak, et al.\cite{Jirak85} at about 400 K.  The data also suggest
a small decrease in width for temperatures greater than $\sim$ 400
K.  Unfortunately, the high temperature limit of our oven
prevented exploring the possible loss of octahedral tilt ordering
above 900 K in this sample.  Moreover, the fact that the
scattering at the orbital wavevector does not completely disappear
makes direct comparisons with the temperature dependence of
LaMnO$_3$ difficult. It remains to fully understand the phase
diagram of PCMO with x=0.25.  Quantitative conclusions will
require further experiments at higher temperatures, including a
search for the octahedral tilt ordering transition.

Similar studies of the orbital and tilt scattering were carried
out in Pr$_{0.6}$Ca$_{0.4}$MnO$_3$ at selected temperatures
between 200 and 300 K.  There is a discontinuous change in the
lattice constants associated with the loss of orbital ordering at
T$_{oo}$=245 K, but neither a decrease in intensity nor a
broadening of the resonant peaks associated with the tilt ordering
was observed. As with PCMO (x=0.25), we were not able to reach the
octahedral tilt ordering transition at high temperatures, and are
again unable to make quantitative comparisons with LaMnO$_3$.
Additional experiments on these and other samples will clearly be
required before these intriguing results can be understood,
especially at elevated temperatures.

\section{Conclusions}

We have presented a detailed study of the x-ray resonant
scattering present at the octahedral tilting wavevector of
Pr$_{1-x}$Ca$_x$MnO$_3$ for x=0.25 and 0.4 and of LaMnO$_3$ for
incident x-ray energies tuned to the Pr and La L$_I$, L$_{II}$ and
L$_{III}$ absorption edges, respectively.  We show firstly that it
is possible to characterize the structure and temperature
dependence of octahedral tilt ordering in manganites using x-ray
resonant techniques, similarly to recent studies of Mn orbital
ordering.  The energy dependence of the lineshapes is
characterized by a main resonant peak near the inflection point of
the fluorescence in each case, consistent with a dipole
excitation. Specifically, they are 2p$_{3/2} \rightarrow$
5d$_{5/2}$,$_{3/2}$ 2p$_{1/2}$ $\rightarrow$ 5d$_{3/2}$, and 2s
$\rightarrow$ 6p for the L$_{III}$, L$_{II}$ and L$_I$ absorption
edges, respectively. In addition, the L$_{II}$ and L$_{III}$ edge
spectra exhibit fine structure peaks about 30 eV higher in energy.
The polarization dependence of the resonant tilt scattering is
found to be $(\sigma-\pi)$ to within the accuracy of our
measurements, while the azimuthal dependence exhibits a
sin-squared variation. Both are consistent with the expectations
of Templeton scattering at this wavevector and for this space
group.  We interpret these results in terms of a simplified model
of the resonant cross-section in which the resonant d and p states
are split by the local crystal field in analogy to a model
proposed for Mn orbital ordering. More accurate band structure
calculations of the cross-section of resonant scattering from
octahedral tilt ordering by Benedetti, \textit{et
al.}\cite{Benedetti} reproduce qualitative features of the
experimental results, including the main dipole resonance and fine
structure, together with the polarization and azimuthal
dependence.  However, we are not able to observe any modification
of the lineshape upon orbital ordering, in contrast to theoretical
predictions.

On the basis of our temperature dependent studies, it was possible
to associate the disappearance of the orbital and tilt scattering
at higher temperatures with known structural transitions.
Intriguingly, we find that the octahedral tilting in LaMnO$_3$ is
resolution-limited (corresponding to domains in excess of 1000
$\AA$) in the presence of orbital ordering, but gradually
disorders beginning at temperatures about 70 K above the orbital
order transition, until octahedral tilt order is lost at 1000 K.
Similar behavior was not observed in the Pr-based materials, at
least over the limited range of temperature available with the
present apparatus. Further studies at higher temperatures nearer
the octahedral tilt order transition are required before direct
comparisons with LaMnO$_3$ are possible. Taken together these
results show that it is possible to study octahedral tilt ordering
in perovskites using x-ray resonant scattering techniques, and
they raise the question of the nature of the coupling between tilt
and orbital ordering.

\section{Acknowledgements}

We acknowledge helpful conversations with M. Blume, G.A. Sawatzky,
T. Vogt and P. Woodward.  The work at Brookhaven, both in the
Physics Department and at the NSLS, was supported by the U.S.
Department of Energy, Division of Materials Science, under
Contract No. DE-AC02-98CH10886.  Support from the Ministry of
Education, Science and Culture, Japan, by the New Energy and
Industrial Technology Development Organization (NEDO), and by the
Core Research for Evolution Science and Technology (CREST) is also
acknowledged. Work at the CMC beamlines is supported, in part, by
the Office of Basic Energy Sciences of the U.S. Department of
Energy and by the National Science Foundation, Division of
Materials Research.  Use of the Advanced Photon Source was
supported by the Office of Basic Energy Sciences of the U.S.
Department of Energy under Contract No. W-31-109-Eng-38.


\begin{thebibliography}{10}
\bibitem{Glazer}
In the Glazer notation\cite{Glazer72}, this structure is obtained by rotations
  about the cubic axes with a pattern a$^{+}$b$^-$b$^-$, that is, one axis
  distinguished from the other two by having in phase rotations of the
  octahedra along that axis. In $\it{Pbnm}$ notation, this axis is the c-axis.
  The net effect of these rotations is mostly a rotation about the orthorhombic
  b-axis, fig. 1  .
\bibitem{fullwidths}
The measured full widths at the L$_{II}$ absorption edge were 6 eV for the PCMO
  (x=0.4) and LaMnO$_3$ samples and 7.5 eV for the PCMO (x=0.25) sample. The
  L$_{III}$ full widths were 6 eV for the LaMnO$_3$ and 9 eV for the PCMO
  (x=0.25)  .
\end{thebibliography}

\begin{figure}
\caption[]{ (a)  A three-dimensional view of a perovskite
structure (Pr$_{0.6}$Ca$_{0.4}$MnO$_3$) including the octahedral
tilt ordering associated with the GdFeO$_3$-type distortion. (b),
(c), and (d) are projections of (a) along orthorhombic c,b and a
axes, respectively. Small clear (dark) spheres represent oxygen
atoms at the corners (Mn atoms at the center) of the octahedra.
Larger spheres represent the A-site ions, either La, Pr or Ca.}
\end{figure}

\begin{figure}
\caption[]{ Scan of the x-ray intensity measured at the (0,1,0)
wavevector of Pr$_{0.75}$Ca$_{0.25}$MnO$_3$ for incident x-ray
energies between 6.4 and 6.64 keV.  The Pr L$_{II}$ and Mn K edge
absorption energies are marked.  The sample was held at 10 K.}
\end{figure}

\begin{figure}
\caption[]{ Scans of the intensity measured at the (0,1,0)
wavevector of Pr$_{0.75}$Ca$_{0.25}$MnO$_3$ versus incident x-ray
energies near the Pr L$_{III}$, L$_{II}$, and L$_I$ absorption
edges.  These data were obtained with a Ge(111) analyzer and so
combine any $\sigma$- and $\pi$- contributions to the scattering.
The sample was held at room temperature in these scans.}
\end{figure}

\begin{figure}
\caption[]{ Scans of the intensity measured at the (0,1,0)
wavevector of LaMnO$_3$ versus incident x-ray energies near the La
L$_{III}$, L $_{II}$, and L$_I$ absorption edges.  Solid circles
show energy scans of the background intensity taken at (0.975, 0,
0.033) for comparison. These data were obtained with a Ge(111)
analyzer and so combine any $\sigma$- and $\pi$- contributions to
the scattering.  The sample also was held at room temperature.}
\end{figure}

\begin{figure}
\caption[]{ Azimuthal dependence of the scattering at the (0,1,0)
wavevector of Pr$_{0.75}$Ca$_{0.25}$MnO$_3$ for incident x-ray
energies at the Pr L$_{II}$ absorption edge.  The solid line is a
fit to the form ASM${^2\psi}$.}
\end{figure}

\begin{figure}
\caption[]{ Intensity of the $\sigma \rightarrow \pi$ component of
the resonant scattering obtained at the (0,1,0) wavevector of
Pr$_{0.6}$Ca$_{0.4}$MnO$_3$ for incident x-ray energies near the
Pr L$_I$ and L$_{II}$ absorption edges.  Open and closed circles
represent data at temperatures above and below the orbital
ordering temperature of T$_{oo}$ = 245 K.}
\end{figure}

\begin{figure}
\caption[]{ (a)  Temperature dependence of the resonant scattering
obtained at the (1,0,0) wavevector of LaMnO$_3$ at the La L$_I$
(open circles), and Mn K (closed circles) absorption edges. (b)
Corresponding plot of the half-widths-at-half-maxima plotted on a
log scale.}
\end{figure}

\begin{figure}
\caption[]{ (a)  Temperature dependence of the resonant scattering
obtained at the (0,1,0) wavevector of
Pr$_{0.75}$Ca$_{0.25}$MnO$_3$ at the Pr L$_{II}$ (open circles)
and Mn K (closed circles) absorption edges.  (b)  Corresponding
plot of the half-width-at-half-maxima.}
\end{figure}

\end{document}